\documentclass[notitlepage,a4paper,aps,prd,onecolumn,superscriptaddress,nofootinbib,groupedaddress]{revtex4}

\usepackage{enumerate}
\usepackage{amsmath}
\usepackage{amsfonts}
\usepackage{amssymb}
\usepackage[utf8]{inputenc}
\usepackage[T1]{fontenc}
\usepackage{mathtools}
\usepackage{wasysym}
\usepackage{accents}
\usepackage{graphicx}
\usepackage[normalem]{ulem}
\usepackage[svgnames]{xcolor}
\usepackage[colorlinks=true]{hyperref}
\usepackage{url}
\usepackage{float}
\usepackage{orcidlink}

\newtheorem{definition}{Definition}

\begin{document}
\title{Muon accelerators - Muon lifetime measurements as window to Planck scale physics}

\author{Iarley P. Lobo\,\orcidlink{0000-0002-1055-407X}}
\email{lobofisica@gmail.com,  iarley_lobo@fisica.ufpb.br}
\affiliation{Department of Chemistry and Physics, Federal University of Para\'iba, Rodovia BR 079 - km 12, 58397-000 Areia-PB,  Brazil}
\affiliation{Physics Department, Federal University of Lavras, Caixa Postal 3037, 37200-900 Lavras-MG, Brazil}

\author{Christian Pfeifer\,\orcidlink{0000-0002-1712-6860}}
\email{christian.pfeifer@zarm.uni-bremen.de}
\affiliation{ZARM, University of Bremen, 28359 Bremen, Germany.}

\begin{abstract}
A prominent effective description of particles interacting with the quantum properties of gravity is through modifications of the general relativistic dispersion relation. Such modified dispersion relations lead to modifications in the relativistic time dilation. A perfect probe for this effect, which goes with the particle energy cubed $E^3$ over the quantum gravity scale $E_{\text{QG}}$ and the square of the particle mass $M^2$ would be a very light unstable particle for which one can detect the lifetime in the laboratory (lab) as a function of its energy measured in the lab to very high precision. In this article we conjecture that a muon collider or accelerator would be a perfect tool to investigate the existence of an anomalous time dilation, and with it the fundamental structure of spacetime at the Planck scale.
\end{abstract}

\maketitle


\section{Introduction}
    A self consistent theory of quantum gravity is still elusive and the theoretical predictions, as well as experimental searches for traces of quantum gravity are part of the important ongoing endeavours in fundamental physics \cite{bookgiulini,Amelino-Camelia:2008aez,bookhoss,Addazi:2021xuf}. Experimentally, so far no clear unambiguous signs of quantum gravity have been found, while on theoretical side there exist several approaches to construct a self-consistent theory of quantum gravity, among the most famous and most studied are canonical quantum gravity \cite{Kiefer:2004xyv}, string theory \cite{Polchinski:1998rq,Polchinski:1998rr}, loop quantum gravity \cite{Ashtekar:2021kfp}, causal dynamical triangulation \cite{Loll:2019rdj} and asymptotic safety \cite{Eichhorn:2018yfc}.

    The main caveat to find quantum gravity effects is that these, if they become relevant at the Planck scale, are highly suppressed: in terms of length by $\ell_{\text{Pl}}\sim 10^{-35}\, \text{m}$ or, in terms of energy by $E_{\text{Pl}}\approx 1.2\times 10^{19}\, \text{GeV}$. Until now, no unambiguous quantum gravity effects have been detected, and the chances are low to detect them by accident. Therefore, to increase the chances for a detection, mainly two things are needed: First, a promising, rigorously derived prediction from a fundamental or phenomenological model of quantum gravity, in order to know where to look for the effect; second, an amplification mechanism which brings the predicted effect into the range of nowadays measurement precision.

    The most prominent amplifiers in the literature appear in the context of cosmic astrophysical or cosmological systems. In the search for quantum gravity induced time delays of high energetic particles (photons or neutrinos), possible tiny modified dispersion relation (MDR) effects could accumulate over the enormous cosmological travel distance of the particles \cite{Pfeifer:2018pty}, and become visible in gamma-ray and neutrino telescopes \cite{FermiGBMLAT:2009nfe,MAGIC:2020egb,IceCube:2021tdn,Amelino-Camelia:2020bvx,Amelino-Camelia:2016ohi}. For quantum gravity modified particle interactions and threshold effects, the amplifier is the power of the energy of the parent particle. In ultra-high energetic cosmic rays the highest energetic particles could trigger such modified interactions that become visible in the observation of the rays and the products of interactions of the rays with the particles of the atmosphere \cite{Jacobson:2002hd,HAWC:2019gui}. Imprints of quantum gravity may also be present in gravitational waves (GWs) signals, for instance, in the profile of GW background, propagation speed of GWs and the luminosity distance of GW sources  \cite{Calcagni:2020ume}, besides on black holes shadows and perturbations, among other observables \cite{Jusufi:2020wmp,Liu:2020ola,Haroon:2017opl,Laanemets:2022rmn}.

    Despite these high-energy, large distance amplifiers in cosmic systems, there also exist, maybe a bit unexpectedly, amplifiers, which open up a window to Planck scale physics in Earth bound, local physical systems. Prominent examples here are cold atoms experiments \cite{Amelino-Camelia:2009wvc,Haine:2018bwu}, or manifestations of generalized uncertainty principles and minimal length scenarios \cite{Hossenfelder:2012jw,Bosso:2023aht,Wagner:2023fmb,Girdhar:2020kfl}.

    Recently \cite{Lobo:2020qoa}, we found an unexpected amplifier in the relativistic lifetime of elementary particles. We investigated how modified dispersion relations, lead to a modified time dilation factor between the laboratory frame and the rest frame of the particle, by applying the clock postulate rigorously to Planck scale modified dispersion relations. The result was, that for certain modified dispersion relations, the time dilation contains a modification which scales with the energy of the particle to the third power, which would lead to a detectable modification for high energetic particles, even if the effect is suppressed by the Planck energy scale. In this letter we outline how such a modified time dilation factor could in principle be detected (or excluded) in dedicated accelerator experiments on Earth.

    The advantage of Earth, or Earth orbit, based experiments over the observation of cosmic messengers is that the setup and the initial conditions of the system under observation can be highly controlled. Moreover, usually the measurement precision and measurement time is usually higher. The downside is that the energies which can be achieved on Earth, or Earth orbit, are not as high as the ones of cosmic messengers.
\section{Time dilation from modified dispersion relations}
    The time dilation effect in Special Relativity is a consequence of any of two features: 
    \begin{enumerate}[a)]
        \item the Lorentz transformation between frames,
        \item the so called  clock postulate, which states that the proper time that an observer measures between two events on spacetime is given by the length of its trajectory between the two events.
    \end{enumerate}
    This second approach has a conceptual advantage in comparison to the former, as it does not rely on Lorentz transformations as symmetry of spacetime and can thus be applied to curved spacetimes as well and includes gravitational time dilations or redshifts.

    Moreover, the clock postulate can be extended, even to spacetime geometries beyond Riemannian spacetime geometry, which makes this approach most suitable to study time dilations in the absence of a spacetime metric. The only geometric ingredient needed on spacetime to employ the clock postulate is a geometric length measure for curves. The most general spacetimes with this property are so called Finsler spacetimes \cite{Pfeifer:2019wus,Albuquerque:2023icp}. Hence, to determine the time dilation induced by Planck scale modified dispersion relations, we need to derive the corresponding length measure/clock for massive particle trajectories.

\subsection{The modified time dilation formula}

    In \cite{Lobo:2020qoa} we performed this derivation in all detail for general modifications of the general relativistic dispersion relation. 
    
\subsubsection{Dispersion relations and Hamilton functions}
    Planck-scale modified dispersion relations can describe departures of the usual kinematics of particles in a quantum spacetime in an effective way. Assuming isotropy, i.e. that the dispersion relation only depends on the norm of the spatial momentum of the particle, $p=|\vec p|$, the MDR can be displayed as set of constant value of a Hamilton function $H(E,p)$,
    \begin{align}\label{eq:modDisp}
        M^2 = E^2 - p^2 + \frac{1}{E_{\text{QG}}} h(E,p)=H(E,p)\,,
    \end{align}
    where $E$ is the energy of the particle, $E_{\text{QG}}$ is the energy scale from which quantum gravity effects become apparent. It suppresses deviations from special/general relativistic expressions. We consider it as proportional to the Planck energy $E_{\text{QG}}=\xi E_{\text{Pl}}$. This assumption is useful to estimate the order of magnitude of the influence of modified dispersion relation, it is however not necessary. In the end experiments will tell us about the value of $E_{\text{QG}}$\footnote{When we can constrain the parameter $\xi$ with bounds of the order 1, we say that the analysis is being done with Planck scale sensitivity \cite{Amelino-Camelia:2009wvc}.}.

    The function $h(E,p)$ depends on the particle's energy and momentum, which for purely polynomial corrections to order $1/E_{\text{QG}}$ in $p_\mu=(E,p)$ would be most generally given by $h(E,p)=\sum_{m=0}^{3}a_{m}p^{m}E^{3-m}$. The choice of the dimensionless parameters $a_{m}$ defines the model under consideration. Eventually, they need to be constrained experimentally or be determined from an underlying theory. Some notable combinations, that are discussed in the literature are: 
    \begin{itemize}
        \item D-brane recoil \cite{Ellis:1999sd}, D-brane foam \cite{Ellis:1999uh} and the bicrossproduct basis of $\kappa$-Poincar\'e \cite{Majid:1994cy}, where the only non-vanishing parameter is $a_2=1$ leading to $h = Ep^2$;
        \item Lioville string QG \cite{Amelino-Camelia:1996bln}, where the only non-vanishing parameter is $a_3=1$ leading to $h = p^3$;
        \item $\kappa$-Poincar\'e in the Magueijo-Smolin basis \cite{Magueijo:2001cr,Kowalski-Glikman:2002iba}, where the only non-vanishing parameters are $a_0=1$ and $a_2=-1$ leading to $h = (E^2-p^2)E$.
    \end{itemize}
    The form of $h(E,p)$ determines if the light cone of massless particles, $H(E,p)=0$, will be more wide of more narrow than the one of special relativity. The former case corresponds to what is called superluminal propagation of massless particles, and the latter corresponds to what is called subluminal propagation. Sub- and superluminal are meant here compared to special relativity. For example, for a correction of the kind $h(E,p)=a_{3}p^3$, if $a_{3}$ is positive (negative), we have subluminal (superluminal) propagation. Both cases are investigated on observational grounds using astroparticle physics techniques \cite{Addazi:2021xuf}. For massive particles the normalization condition (the mass shell) $H(E,p)=m^2$ is deformed away from the usual relativistic hyperboloid, which ist he source of the modified time dilation effects, that we discuss below.

\subsubsection{Time measurement: the induced length measure for worldlines and time dilation}
    The action that gives the equations of motion of a particle subject to a MDR is the so called Helmholtz action \eqref{eq:modDisp} and is given by the functional
    \begin{equation}
        S[x,p,\lambda]=\int d\sigma[\dot{x}^{\mu}p_{\mu}-\lambda(H-M^2)]\, ,
    \end{equation}
    where $x^{\mu}(\sigma)=(t(\sigma),\vec{x}(\sigma))$, $p_{\mu}(\sigma)=(E(\sigma),\vec{p}(\sigma))$, $\lambda$ is a Lagrange multiplier and ``dot'' means derivative with respect to the parameter $\sigma$. As discussed in earlier papers, e.g. \cite{Girelli:2006fw,Amelino-Camelia:2014rga,Lobo:2016xzq,Lobo:2020qoa,Pfeifer:2019wus,Rodrigues:2022mfj}, it is equivalent to an action in the Lagrangian formalism, which can be expressed as
    \begin{equation}\label{eq:s}
        s[x]=\int F(x,\dot{x})d\sigma\, ,
    \end{equation}
    where the function $F$ is a $1$-homogeneous function with respect to $\dot x$ and thus $s[x]$ is independent of the pramaterization of the curve $x(\sigma)$. This makes $s[x]$ a geometric quantity, which only depends on the curve. In special and general relativity we would find that $F(x, \dot x) = \sqrt{g_{\mu\nu}(x)\dot x^\mu \dot x^\nu}$. The further geometric interpretation of this formalism is achieved by identifying the action \eqref{eq:s} with the arc-length functional that defines a Finsler function $F$ in spacetime. For this reason, we can state that {\it Finsler spacetime geometry} (${\cal M},F)$ on a manifold ${\cal M}$ describes the kinematics of particles subject to a MDR.

    In Finsler geometry, the geometry of spacetime is derived from $F$ as fundamental variable in a similar way as the geometry of spacetime is derived in special and general relativity from a spacetime metric $g$. It leads to a Finsler metric (a velocity dependent metric) that is defined as the Hessian of the square of the Finsler function \cite{Miron:1994nvt} (which is the metric of the effective Finslerian quantum spacetime), 
    \begin{align}
        g_{\mu\nu}(x,\dot x) = \frac{1}{2}\frac{\partial^2}{\partial \dot x^{\mu} \partial \dot x^{\nu}}F^2(x,\dot x)\,.
    \end{align}
    Furthermore, on Finsler spacetimes, a causal structure with null, timelike and spacelike curves is well defined \cite{Minguzzi:2014aua,Javaloyes:2018lex,Hohmann:2021zbt}, geodesics extremize the arc-length functional, are the worldlines of point particles and are equivalent to the solutions of the Hamiltonian equations of motion of $H$, the mass shell is simply the hypersurface found from the norm of the conjugate four-momentum and the isometries are transformations that preserve the form of the arc-length functional \cite{Girelli:2006fw,Amelino-Camelia:2014rga,Lobo:2016xzq}. The latter property is of particular interest to the Doubly Special Relativity (DSR) community \cite{Amelino-Camelia:2000stu}, since it implies in the possibility of deforming rather than breaking Lorentz symmetry despite the presence of a MDR (we shall further discuss this issue later in this paper).

    The action \eqref{eq:s} also allows one to extend the clock postulate in this effective spacetime in the following way:
    \begin{definition}[Clock Postulate]
    The proper time an observer, or massive particle, experiences between events $A$ and $B$ along a timelike curve (her worldline) in a Finsler spacetime $({\cal M},F)$ is the length of this curve between events $A$ and $B$
    \begin{equation}
        \tau=\int_{\sigma_A}^{\sigma_B}F(x,\dot{x})d\sigma\, .
    \end{equation}
    \end{definition}
    In \cite{Lobo:2020qoa} it was shown that if the MDR \eqref{eq:modDisp} only depends on $p$, as we assume in this study, then the Finsler function $F(x,\dot{x})$ only depends on $v=|\vec v|$, where $\vec v = d\vec x/dt$ and $t$ is the laboratory time parameter. For such spatially isotropic proper time measures, we can write down a relation between the proper time measured by the particle, $\tau$, and the laboratory time, $t$, as
    \begin{align}
        \tau = \frac{t}{\gamma} \left( 1 + \frac{1}{E_{QG}} f(\gamma)\right)\,,
    \end{align}
    where $\gamma=(1-v^2)^{-1/2}$ and $f(\gamma)$ is determined from the function $h(E,p)$ which defines the MDR. 
    
   We like to point out here that for circular motion, as it happens in particle colliders, one simply can express $v$ in spherical coordinates $v = \sqrt{v_r^2 + r^2 (v_\theta^2 + \sin^2(\theta) v_\phi^2)}$ and set $\theta =\pi/2$, $v_r=v_\theta = 0$. The angular motion of the particle is then determined by its angular momentum $v_\phi = v_\phi(\mathcal{L}, m, R)$, where $m$ is the particle mass and $R$ the radius of the circular motion of the particle. The precise functional form of $v_\phi(\mathcal{L}, m, R)$ is determined from the Euler-Lagrange equations for $\phi$; the angular momentum is $\mathcal{L} = \partial F^2/\partial \dot \phi$, as usual in classical Lagrangian mechanics. Here, the Lagrangian is given by $L=F^2$. Hence, for circular motion, all the derivations below in terms of the relativistic $\gamma$-factor are applicable, only that in circular motion $\gamma$ is determined by the angular velocity $v_\phi(\mathcal{L}, m, R)$ alone.

\subsubsection{Time dilation in the $\kappa$-Poincar\'e model}
    One of the most studied model of a deformation of Lorentz symmetry is the $\kappa$-Poincar\'e algebra \cite{Lukierski:1991pn,Lukierski:1992dt,Majid:1994cy}. It introduces a quantum deformation of the Poincar\'e algebra by a parameter with dimension of energy, called $\kappa$ (which we treat as the quantum gravity energy scale
    \footnote{The relation between quantum gravity and deformation, rather than violation, of Lorentz symmetry has been shown, for example, in the cases of  $2+1$ dimensional quantum gravity \cite{freidellivineeffective,MatschullWelling}, which is the only QG proposal that has been exactly solved. It shows that the momentum space of a single particle, when gravity is weak, would present anti-de Sitter geometry, thus deforming the Poincar\'e symmetry. Also, from linearization of the polymeric-modified Hamiltonian constraints of general relativity inspired by loop quantum gravity techniques \cite{Amelino-Camelia:2016gfx}, leads to a deformation of the Poincar\'e algebra of symmetry  generators.}, 
    therefore $\kappa=E_{\text{QG}}$), that reduces to the undeformed Poincar\'e algebra when $\kappa$ is much larger than the generators of time and space translations (the four-momenta). Sometime the algebra is also interpreted as introducing a non-commutativity in spacetime, since one version of a $\kappa$-deformation leads to non-commuting generators of translations \cite{Ballesteros:2016bml}. A spacetime whose symmetry algebra is given by the $\kappa$-Poincar\'e algebra is called $\kappa$-Minkowski spacetime.
    
    As the Poincar\'e algebra, the $\kappa$-Poincar\'e algebra possesses two Casimir operators (operators which commute with all other operators). One is the mass operator, which is identified with the Hamilton function defining the modified dispersion relation, the other one is the spin operator. They assume different forms in different momentum space bases. One of the most appealing, from the observational point of view, of these bases is the bicrossproduct one \cite{Majid:1994cy}, which in a first order approximation presents a MDR (mass Casimir) of the kind given by Eq.\eqref{eq:modDisp}, where $h(E,p) = - E p^2$. The Finsler function derived from this MDR in Cartesian coordinates is given by
    
    \begin{align}\label{eq:finslerf}
        F(\dot{t},\dot{\vec{x}})
        &=\sqrt{\dot{t}^2-|\dot{\vec{x}}|^2}+\frac{M}{2E_{\text{QG}}}\frac{|\dot{\vec{x}}|^2\dot{t}}{\dot{t}^2-|\dot{\vec{x}}|^2}\\
        &=\dot t \left(\sqrt{1-v^2}+\frac{M}{2E_{\text{QG}}}\frac{v^2}{1-v^2}\right)\,.
    \end{align}
    The resulting time dilation, expressed in terms of the energy $E$ of the particle, contains a term $E^3$
    \begin{equation}
        t(E,M) 
        = \gamma \tau \left(1 - \frac{M}{2 E_{\text{QG}}}\gamma(\gamma^2 - 1)\right) 
        = t_{\text{SR}} \left( 1 + \frac{M}{2 E_{\text{QG}}} \left( \frac{M}{E} - 2 \frac{E}{M} + \frac{E^3}{M^3} \right)  \right)\,,\label{dil-lt}
    \end{equation}
    where $t_{\text{SR}}=E\tau/M$ is the result from Special Relativity. All details on the derivation of this result can be found in~\cite{Lobo:2020qoa}.
    
    Thus, even if $E_{\text{QG}}$ is the Planck energy, measuring the lifetime of particles of reasonably high energy $E$ can lead to an observable deviation from the special relativistic prediction. Such an amplification effect does not only emerge in this example MDR, but for many models with polynomial modifications of the speical relativistic dispersion relation.

    Therefore, we conjecture to devise a dedicated experiment to measure $t(E,M)$ for unstable particles, since these measurements are a window to new physics and capable to constrain, or find evidence for, deviations from local Lorentz invariance. The special relativistic term, the linear term in $E$ of $t(E,M)$ might just be the first order approximation, to a more complex dependency of the dilated lifetime of particles on their energy. Notice that \eqref{dil-lt} represents a map between the comoving (proper) time and the laboratory time. As we shall later discuss, this map is actually a deformed Lorentz transformation and represents the time-part of an isometry given by a deformed boost between the comoving and the laboratory frames.

\subsection{The actual observable}
    In actual experiments like accelerators, one usually does not consider $t(E,M)$, when one analyses the lifetime of fundamental particles. The quantity considered is the the lifetime of the particle at rest $\tau$ as a function of its mass defined by the particle data group \cite{Workman:2022ynf} $M_{\text{PDG}}$, its transverse momentum $p_{\text{T}}$ and the decay distance of the $L_{xy}$ in the laboratory, \cite{ATLAS:2012cvl,ALICE:2023ecf},
    \begin{align}
        \tau = \tau(M_{\text{PDG}}, L_{xy}, p_\text{T})\,.
    \end{align}
    We consider a simplifyied situation of motion in 1+1 D, such that for the modified dispersion relations, we identify $M_{\text{PDG}}=M$ and $p_{\text{T}}= p = |\vec p|$, so that the only variable left to be identified is $L_{xy}$. 

    Before continuing, we like to remark that for circular motion on a circular path with fixed radius $r=R$ and angle $\theta=\pi/2$, we have that $p = \sqrt{p_r^2+ r^{-2}(p_\theta +\sin^{-2}(\theta) p_\phi^2)} = R p_\phi$, where $p_\phi$ is the angular momentum. Using the replacements $x(\tau) \to \phi(\tau)$ and $p=R p_\phi$ , the results below easily carry over to the case of circular motion. For such a replacement one neglects the details about a force, which keeps the particle on the circular trajectory $r=R$, one simply assumes the particle is fixed on a trajectory $(r=R, \theta=\pi/2, \phi=\phi(\tau))$ and studies the circular motion in $\phi$ direction.

    Considering \eqref{eq:modDisp} as Hamilton function which determines the motion of particles satisfying $M = H(x,p)$. The Hamilton equations of motion lead to energy and momentum conservation, since we are considering translation invariant dispersion relations $H(x,p)=H(p)$, 
    \begin{align}
        \dot E = -\partial_t H = 0\,,\quad
        \dot p = -\partial_x H = 0\,,
    \end{align}
    and determine the worldlines of the particles, here for the $\kappa$-Poincar\'e model $h(E,p) = - E p^2$,
    \begin{align}
        \dot t = \partial_E H  = 2 E - \frac{1}{E_{\text{QG}}} p^2\,, \quad
        \dot x = \partial_{p} H = - 2 p \left(1 + \frac{1}{E_{\text{QG}}} E \right)\,.
    \end{align}
    Hence for the motion of the particle in the lab frame we find, using the dispersion relation $M^2 = E^2 - p^2 - \frac{1}{E_{\text{QG}}} E p^2$ to replace $E$ in terms of $p$ and $M$,
    \begin{align}\label{eq:v}
      v &= \frac{dx}{dt} =\frac{\dot x}{\dot t}
         = -\frac{p}{\sqrt{M^2+p^2}}-\frac{p}{E_{\text{QG}}}\Rightarrow\quad
        x(t)=x(0) - t\, p\left(\frac{1}{\sqrt{M^2+p^2}}+ \frac{1}{E_{\text{QG}}} \right)\, .
    \end{align}
    Fixing the boundary conditions such that the particle gets created at $x(0)$ and decays at $x(t)$, we identify its decay length as $L_{xy} = |x(t)-x(0)|$. Using \eqref{dil-lt} to express the decay length as function of the proper lifetime $\tau$ of the particle and solving for $\tau$ yields
    \begin{align}\label{eq:tau(Lmp)}
        \tau = \frac{L_{xy} M}{p}\left(1 - \frac{1}{E_{\text{QG}}} \frac{\sqrt{M^2+p^2}(2 M^2 + p^2)}{2 M^2 }\right)\approx \frac{L_{xy} M}{p}\left(1 - \frac{1}{E_{\text{QG}}} \frac{p^3}{2 M^2 }\right)\,.
    \end{align}

Notice that actually, this relation between the decay length and the proper lifetime along with the modified time dilation is a deformed Lorentz transformation in velocity space, i.e., an isometry of the Finsler function. This can be verified by using Eqs.\eqref{dil-lt} and \eqref{eq:tau(Lmp)} to write $\dot{t}$ and $\dot{x}$ as functions of the derivative of the proper time $\dot{\tau}$ (which can easily be done since $E$ and $p$ are constants of motion $\dot{E}=0=\dot{p}$)
\begin{align}
\dot{t}&=\dot{\tau}\frac{E}{M} \left( 1 + \frac{M}{2 E_{QG}} \left( \frac{M}{E} - 2 \frac{E}{M} + \frac{E^3}{M^3} \right)  \right)\label{eq:tdot}\\
\dot{x}&=\dot{\tau}\frac{p}{M}\left(1 + \frac{1}{E_{\text{QG}}} \frac{\sqrt{M^2+p^2}(2 M^2 + p^2)}{2 M^2 }\right)\, .\label{eq:xdot}
\end{align}
From these expressions \eqref{eq:tdot} and \eqref{eq:xdot}, we can plug into the Finsler function \eqref{eq:finslerf} to find (for $\dot{\tau}>0$)
\begin{equation}
 F(\dot{t},\dot{x})=\sqrt{\dot{t}^2-\dot{x}^2}+\frac{M}{2E_{\text{QG}}}\frac{\dot{x}^2\dot{t}}{\dot{t}^2-\dot{x}^2}=\dot{\tau}=F(\dot{\tau},0)\, .
\end{equation}
So, as can be seen, this map preserves the Finsler function in the lab and comoving frames. This shows that the time dilation from the clock postulate can also be found from a deformed Lorentz transformation between the comoving and the lab frames. For this reason, just like we did for the clock postulate, we set this property in the form of a definition:
\begin{definition}[Deformed Lorentz Transformation]
    A coordinate transformation that preserves the form of Finsler function is called an isometry. In particular, an isometry that reduces to Lorentz transformations in the Minkowskian limit is called a Deformed Lorentz Transformation.
\end{definition}
Deformed Lorentz transformations from the comoving to the lab frame acting on the momenta can be obtained from the Finsler function, by setting
\begin{align}
    E'(v) = \frac{\partial F}{\partial \dot x^0}\,,\quad p_i'(v) = \frac{\partial F}{\partial \dot x^i}\,.
\end{align}
They satisfy that $H(E'(v),\vec p'(v)) = M^2$ for all $v$, as has been discussed and demonstrated on explicit examples in \cite{Lobo:2020qoa}, and been extended to deformed Lorentz transformations between general frames in \cite{Lobo:2021yem} and \cite{Morais:2023amp}.

Endowed with these results, our prediction is that if the lifetime \eqref{eq:tau(Lmp)} is calculated using data from particle accelerators with higher and higher energies, discrepancies should eventually emerge when we compare this result with the lower energies measurements, if the determination of the lifetime in accelerators is done assuming the special relativistic expression $L_{xy} M/p$. We conjecture to perform searches for deviations from Lorentz invariance this way.

\subsection{Towards detecting anomalous time delays in particle decays}
    Our discussion in the previous section suggests that the special relativistic time dilation might just be the first order of a Taylor series of a more complicated time dilation mechanism, whose coefficients could be determined measuring $t(E,M)$ to high precision. Any deviation from the linear relation would indicate a modification of local Lorentz invariance. 

    The question is, how could one measure the time dilation to the required precision? What would be good search strategies? In the following we discuss some ideas, with the aim to start a discussion about a dedicated measurement of $t(E,M)$.

\subsubsection{Measuring $t(E,M)$ in different experiments, which are working at different energy scales.}
    This strategy is similar, in a certain sense, to the one that has been carried out recently concerning time delays of photons emitted from gamma ray bursts (GRBs) \cite{Zhang:2014wpb,Amelino-Camelia:2016ohi,Bolmont:2022yad}, where an ensemble of astrophysical events is considered in the same analysis and deviations from a horizontal line (in which there is no time delay) would be a signature of a deformed kinematics of photons propagating in a quantum spacetime.

    Unfortunately, there are some difficulties for this approach: although we are capable of reaching  controllable energies at the TeV scale with current experiments, like those performed at the LHC (and possibly for the future the FCC), this is only valid for the hadron beams. The produced particles have energies (or transverse momenta) distributed along a range of a few GeV \cite{ALICE:2013cdo} (the issue on the necessary energy for reaching Planck scale sensitivity will be discussed below). This is known from dedicated analyses of the distribution of produced particles per energy \cite{ALICE:2010syw} in such experiments, and this range is considered as an input in the likelihood analysis carried out for the determination of particles' lifetimes. Therefore, we do not have the direct information regarding the duo ``unstable-particle dilated lifetime'' and ``unstable-particle energy'' and even if we had it, the energies/momentum involved are some orders of magnitude below the TeV scale of the beam.

\subsubsection{Search for a momentum dependence of $L_{xy} M/p$}
    A closer look on \eqref{eq:tau(Lmp)} shows that we can reinterpret the corrections to the proper time of the particle as a momentum dependence of the quantity $L_{xy} M/p$. Special Relativity predicts that the ratio of the the particle's travel distance until it decays and its momentum always gives a constant number $\tau$, as can be seen in \eqref{eq:tau(Lmp)} for $E_{\textrm{QG}}\to\infty$. Therefore, if one is able to consider ranges with higher energies and lengths, we should expect a departure of the constancy of this ratio. In fact, if the effect that we are describing in this paper exists, we predict the momentum dependence for the $\kappa$-Poincar\'e dispersion relation to be
    \begin{align}\label{ratio-finsler}
    \frac{L_{xy} M}{p} = \tau \left(1 + \frac{1}{E_{\text{QG}}} \frac{\sqrt{M^2+p^2}(2 M^2 + p^2)}{2 M^2 }\right)\,.
    \end{align}
    Therefore, one should observe a shift in the likelihood fit of such ratio at higher momentum/length ranges. An actual discrepancy would just be actually perceivable when the relative uncertainty of measurements of this quantity matches the dimensionless correction of Eq.~\eqref{ratio-finsler}. 
    We like to point out that for small masses the effect becomes larger, as the correction term diverges for $M\to0$.

    Let us discuss the recent measurement of the lifetime of the $\Lambda$ hyperon performed at ALICE \cite{ALICE:2023ecf} in this context. From this we can get an estimate between the order of magnitude of the momentum of the decaying particle and the magnitude of the uncertainty in the measurement. This particle was chosen not just because its lifetime encompasses a very recent control of uncertainties, but also because it is the lightest hyperon and we see from \eqref{ratio-finsler} that the lighter the particle, the stronger is the effect. 
  
    The $\Lambda$ hyperon has a an average PDG mass of $M_{\Lambda}=1115.683\, \text{MeV}$ \cite{Workman:2022ynf} and two two-body decay channels $\Lambda\rightarrow p+\pi^{-}$ and  $\bar{\Lambda}\rightarrow \bar{p}+\pi^{+}$. Its lifetime has been reconstructed at ALICE as $\tau_{\Lambda}=[261.07\pm 0.37\, \text{(stat.)}\pm 0.72\, \text{(syst.)}]\, \text{ps}$, and the particles were produced from a Pb-Pb beam collision with a center of mass energy of $\sqrt{s_{\text{NN}}}=5.02\, \text{TeV}$. 
    
    This setup gives a relative uncertainty of the order $\sigma\sim 0.1\%$, as can be seen if we express the result of the measurement as $\tau_{\Lambda}=261.07[1\pm 0.14\%\, \text{(stat.)}\pm 0.28\%\, \text{(syst.)}]\, \text{ps}$, and we estimate $\tau_{\Lambda}\sim 261.07[1\pm \sigma]\, \text{ps}$.
    
\begin{figure}[H]
    \centering
    \includegraphics[scale=0.5]{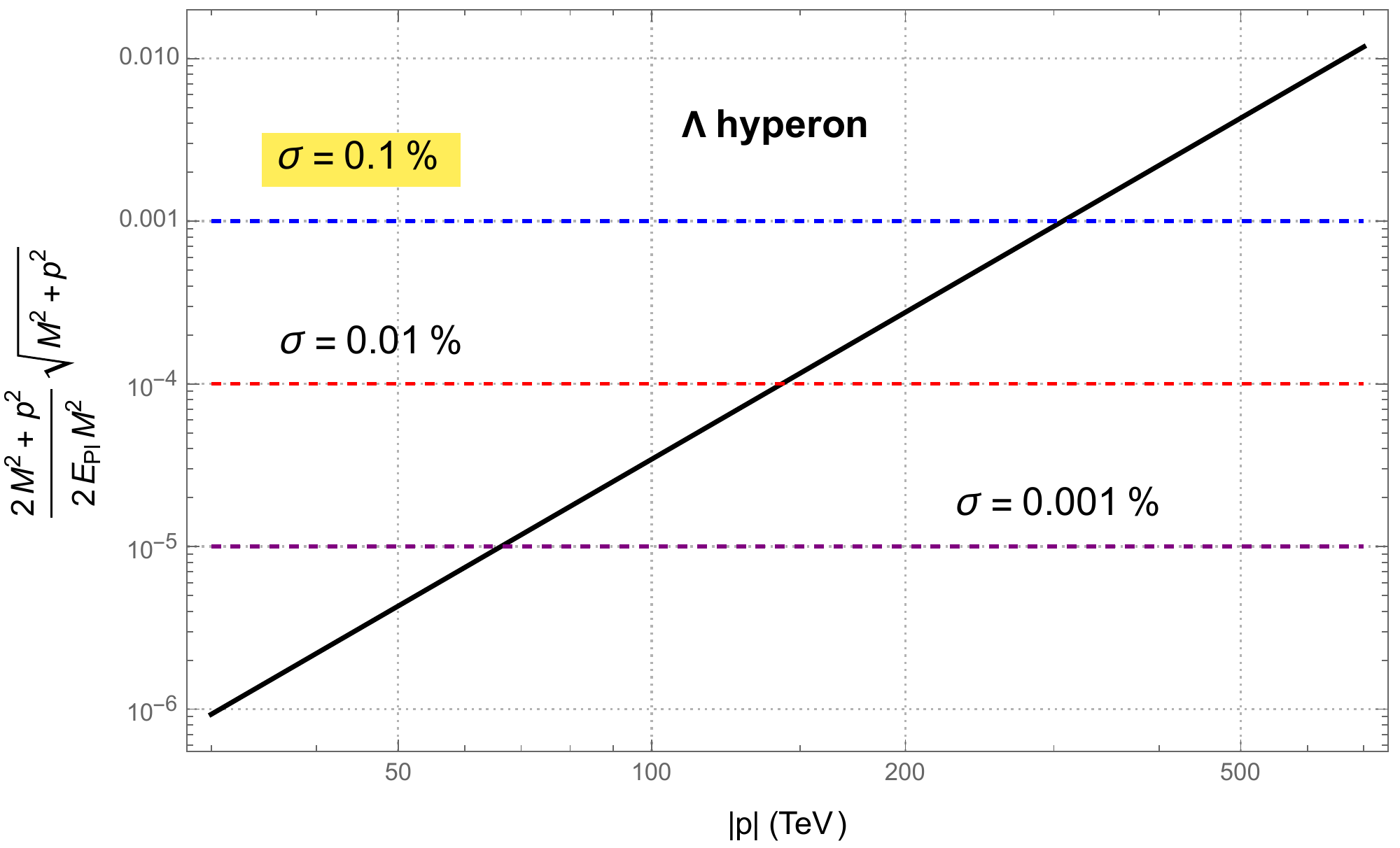}
    \caption{The dimensionless correction of the lifetime as function of the momentum (solid, black line) for $E_{\text{QG}}=E_{\text{Pl}}=1.2\times 10^{19}\, \text{GeV}$ (Planck energy). We considered the $\Lambda$ hyperon mass $M=1115.7\, \text{MeV}$, whose lifetime has been measured in \cite{ALICE:2023ecf} with relative uncertainty $\sigma=0.1\%$ (blue, dashed line), where we see the need for $|p|\sim 300\, \text{TeV}$ to achieve Planck scale sensitivity. We also described hypothetical relative uncertainties of $\sigma=0.01\%$ (red, dashed line) and $\sigma=0.001\%$ (purple, dashed line), where we see the need for $|p|\sim 150\, \text{TeV}$ and $\sim 60\, \text{TeV}$ to achieve Planck scale sensitivity. We highlighted the precision achieved nowadays for an actual measurement of the $\Lambda$ hyperon lifetime.}
    \label{fig:correction-uncertainty-hyperon}
    \end{figure}

    In Fig.\ref{fig:correction-uncertainty-hyperon}, we plotted the dimensionless correction in Eq.\eqref{ratio-finsler} as a function of the momentum, assuming $E_{\text{QG}}=E_{\text{Pl}}=1.2\times 10^{19}\, \text{GeV}$ which the Planck energy. We see that in order to achieve Planck scale sensitivity by performing experiments of the lifetime with nowadays uncertainty of $0.1\%$, we would need a momentum range of the order $300\, \text{TeV}$ (blue, dashed line). Furthermore, even if we somehow managed to improve the measurement precision in one order of magnitude, going to $\sigma\sim 0.01\%$ (red, dashed line), or two orders of magnitude, going to $\sigma\sim 0.001\%$ (purple, dashed line), we would still need a momentum range of the order $150\, \text{TeV}$ or $\sim 60\, \text{TeV}$, respectively. This scenario is unachievable in the foreseen future regarding the energy of the hadron beam (for the first two uncertainties) and even more regarding the momentum range of the produced $\Lambda$ hyperon.

\begin{figure}[H]
    \centering
    \includegraphics[scale=0.5]{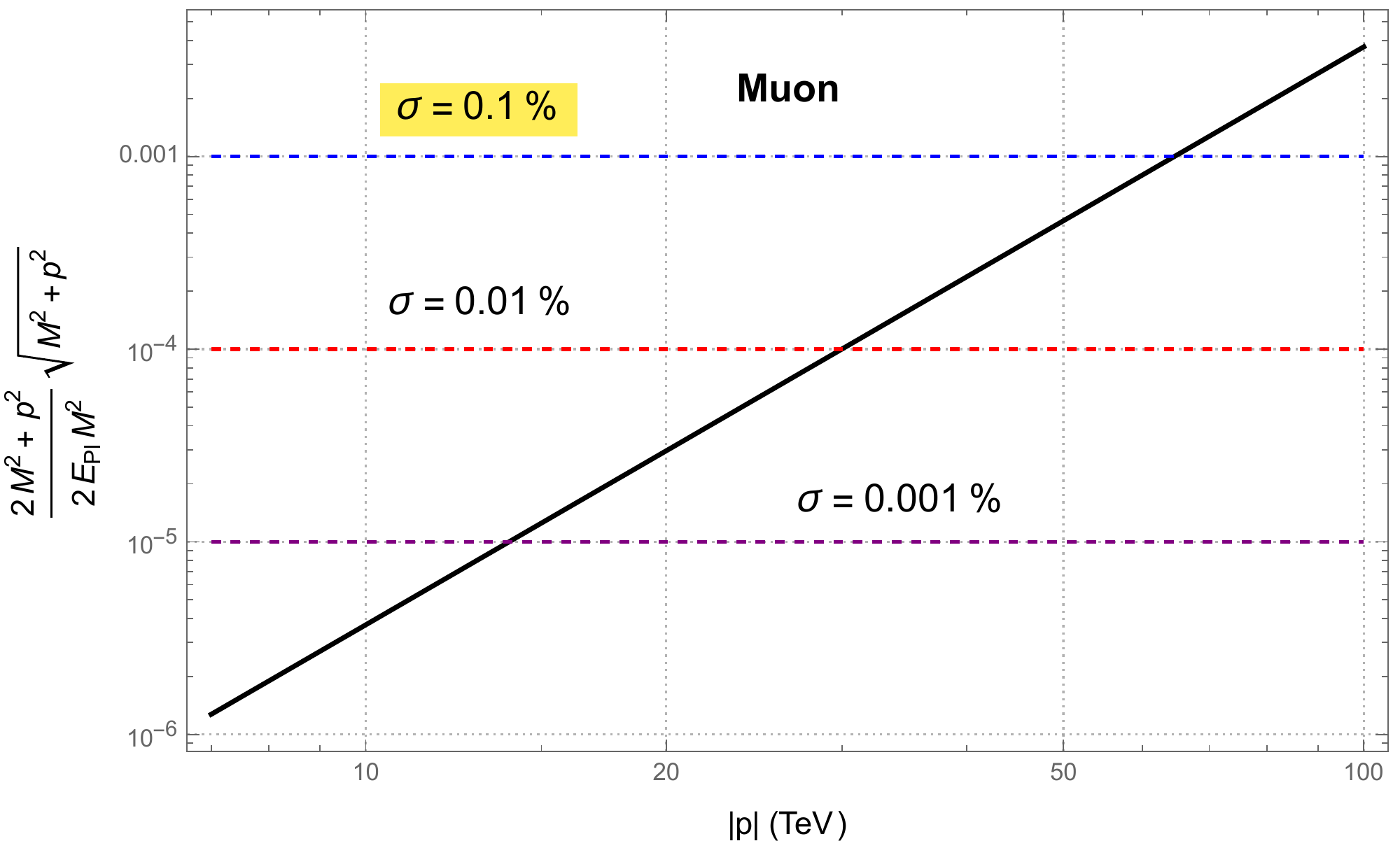}
    \caption{The dimensionless correction of the lifetime as function of the momentum (solid, black line) for $E_{\text{QG}}=E_{\text{Pl}}=1.2\times 10^{19}\, \text{GeV}$ (Planck energy). We considered the muon mass $M=105.7\, \text{MeV}$. The horizontal lines represent some possible relative uncertainties: $\sigma=0.1\%$ (blue, dashed line), where one would need for $|p|\sim 65\, \text{TeV}$; $\sigma=0.01\%$ (red, dashed line), where one would need for $|p|\sim 30\, \text{TeV}$; $\sigma=0.001\%$ (purple, dashed line), where one would need for $|p|\sim 14\, \text{TeV}$ to achieve Planck scale sensitivity. We highlighted nowadays precision in lifetime measurements in accelerators.}
    \label{fig:correction-uncertainty-muon}
    \end{figure}

    As can be seen in \eqref{ratio-finsler}, a way to improve this effect is to consider light particles, however, they cannot be so light that it has a too long dilated propagation distance such that the result of its decay is produced beyond the detector. For example, the experiments carried out at the LHC is uncapable of detecting the product of the decay of the muon, which is a light particle with an average PDG mass $M_{\pi}=105.658\, \text{MeV}$. If the lifetime of the muon could be measured at such accelerators, we would be able to reduce the necessary energy for reaching the Planck scale as can be seen in Fig.\ref{fig:correction-uncertainty-muon}. This figure shows that for nowadays control of uncertainties (blue, dashed line), we would need muons with momenta of the order $65\, \text{TeV}$. Each order of magnitude in improvement in the precision of the measurement reduces in half the necessary momentum for reaching Planck scale sensitivity, where we reach LHC-like energies only for a relative uncertainty of the order $10^{-6}$. In any of these cases, we still have the problem for hadron colliders in which only the hadrons beam would achieve such energies, and the energy of the unstable particle itself would be just a fraction of it, which does not help us in scrutinizing the Planck scale.
    
\subsubsection{Accelerating light unstable particles: muon accelerators}
    The main lesson from the two previous subsections is that it is difficult to use hadron colliders to measure the lifetime, or travel distance of unstable particles to a precision needed to detect Planck-scale induced deviations from Lorentz invariance. 

    What is needed are light unstable particles which can be accelerated to energies which are achievable in hadron accelerators like the LHC or FCC. \vspace{10pt}

    \emph{The most natural candidate for such an undertaking which satisfies this requirement is to study the lifetime of muons and to build a muon accelerator/collider.}\vspace{10pt}

    A muon collider \cite{Long:2020wfp} is a proposal that has recently gained attraction in the community, including some Snowmass papers dedicated to its idealization, summarized in \cite{Black:2022cth} and the establishment of the International Muon Collider Collaboration \cite{Black:2022cth}. The main objective of these efforts is the construction of a $10+$TeV accelerator capable of colliding muons, which would have the advantage of allowing the exploration of a leptonic environment at energies that are higher than those achievable for electron-positron colliders \cite{Black:2022cth}. Besides that, it would provide a cleaner environment for collecting data for such unstable particle in comparison to hadron colliders. The cleanness of the environment is also fundamental to compare bounds found for deviations of special relativity dilated lifetimes from cosmic rays data \cite{PierreAuger:2021mve}. The analysis of particles in extensive air showers heavily suffers from uncertainties \cite{Addazi:2021xuf} such that it becomes hard to find smoking guns that could point out to new physics from the presence of anomalies in this scenario, like explaining the recently found muon deficit \cite{PierreAuger:2021qsd}.

    The idea of having a muon beam produced and accelerated in a ring by a magnetic field and measuring the result of its decay in order to verify its dilated lifetime is not new and has been used in the past to verify the time dilation prediction of Special Relativity \cite{Bailey:1977de} at $0.1\%$ relative uncertainty. In this case, the apparatus that served to verify such dilation was the same as the one used to measure the anomalous magnetic moment of the muon at CERN. 
    
    Although this kind of experiment served well to verify the standard time dilation, it would not be suitable to test Planck scale corrections. For the muon $g-2$ measurements it is necessary to have a fixed and low (for our standards) ``magic'' Lorentz factor $\gamma\approx 29.3$. This is needed in order to remove the contribution of the stabilizing quadrupole electrostatic field from the muon’s relation between the angular frequency and the electromagnetic field according to Thomas-Bargmann-Michel-Telegdi equation \cite{jegerlehner2007anomalous,Lobo:2018zrz}.
    
    We imagine that if a similar scenario could be realized, but for a $10+$TeV muon accelerator in the future, this would lead to fundamental new insights on the validity of Lorentz invariance as fundamental symmetry of nature at these high energy scales.\vspace{10pt}
    
    A most spectacular outcome of such a measurement would be the evidence that indeed Lorentz invariance is just the low energy approximation of a more fundamental symmetry, the less spectacular, but not lesser important outcome would be to constrain models of quantum gravity.

\section{Conclusion}
    Recently, it has been suggested that an underlying quantum spacetime structure can cause deviations in the dilated particle lifetime $t(E)$ of high energetic particles from the special relativistic time dilation. The latter being linear in the particles energy $E$, the quantum spacetime corrections are proportional to $E^3$ for some models, which we discussed here.
    
    In actual measurements, the decay distance $L$ is usually measured. Knowing also the particle's mass $M$ and its momentum $p$ one can obtain the particles rest-frame lifetime $\tau(M,L,p)$. Due to the modified time dilation factor $t(E)$, corrections emerge in $\tau$ of the order of square of the ratio between the energy of the particle and its mass, which serve as amplifier of Planck scale effects. They represent a great opportunity for phenomenological analyses carried out using light and high energetic unstable particles.
    
    Preliminary studies of this effect have been performed in the environment of extensive air showers from cosmic rays, but with a drawback that such environment is very polluted with uncertainties, making it hard to find a smoking gun, i.e., an unambiguous signal that could just be explained by quantum gravity. In order to clean up such environment and aid in the search for such effect, we suggest that particle accelerators could indeed serve to scrutinize this scenario, at least at first order perturbation in the supposedly Planckian energy scale.
    
    In this paper, we demonstrated the existence of the effect, based on modified dispersion relations, and discussed some difficulties that one would face when investigating this effect even at a next generation hadron collider, that could reach energies $50+$TeV, like the FCC. The main difficulty concerns the fact that the produced, unstable particles would have just a fraction of the energy of the primary beam at the order of tens of GeVs. However, as we verified, one would need to reach energies beyond the TeV scale in order to scrutinize this effect with Planck scale sensitivity.
    
    The most prominent solution to this technological problem is actually the development of an accelerator capable of accelerating a light unstable particle to such energy scale. This candidate very naturally turns out to be a {\it muon collider or accelerator} capable of reaching $10+$TeV, that is currently under discussion for the next twenty years. Therefore, this paper adds an extra brick to the set of possible achievements of such apparatus, for the research on quantum gravity and fundamental spacetime symmetries. Moreover, this paper presents a first step and motivation of a further detailed study of the realisation of an experimental setup which can measure the lifetime of particles as function of their energy to high precision.
    
\section*{Acknowledgments}
    I. P. L. was partially supported by the National Council for Scientific and Technological Development - CNPq grant 306414/2020-1 and by the grant 3197/2021, Para\'iba State Research Foundation (FAPESQ). The authors would like to acknowledge networking support by the COST Action QGMM (CA18108), supported by COST (European Cooperation in Science and Technology). 
    C.P. was funded by the cluster of excellence Quantum Frontiers funded by the Deutsche Forschungsgemeinschaft (DFG, German Research Foundation) under Germany's Excellence Strategy - EXC-2123 QuantumFrontiers - 390837967.




\bibliographystyle{utphys}
\bibliography{PLtoPL}

\end{document}